\documentclass[twocolumn]{autart}    

\usepackage{graphicx}          
\usepackage{cite}
\usepackage{amsmath,amssymb,amsfonts}
\usepackage{xcolor}
\usepackage{graphicx}
\usepackage{hyperref}
\usepackage{xurl}
\hypersetup{hidelinks}
\usepackage{textcomp}
\usepackage{physics}
\usepackage{subcaption}

\begin{document}
\baselineskip=.95 \normalbaselineskip 
\allowdisplaybreaks
\begin{frontmatter}

\title{Operator Learning for Robust Stabilization of Linear Markov-Jumping Hyperbolic PDEs} 

\thanks[footnoteinfo]{Corresponding author: Huan Yu (huanyu@hkust-gz.edu.cn)}

\author[1]{Yihuai Zhang}\ead{yzhang169@connect.hkust-gz.edu.cn},    
\author[2]{Jean Auriol}\ead{jean.auriol@centralesupelec.fr},               
\author[1]{Huan Yu}\ead{huanyu@hkust-gz.edu.cn} 

\address[1]{The Hong Kong University of Science and Technology (Guangzhou), Thrust of Intelligent Transportation, Guangzhou, China.}  
\address[2]{Universit\'{e} Paris-Saclay, CNRS, CentraleSup\'{e}lec, Laboratoire des Signaux et Syst\`{e}mes, 91190, Gif-sur-Yvette, France.}             


\begin{abstract}                          
This paper addresses the problem of robust stabilization for linear hyperbolic Partial Differential Equations (PDEs) with Markov-jumping parameter uncertainty. We consider a 2 $\times$ 2 heterogeneous hyperbolic PDE and propose a control law using operator learning and the backstepping method. Specifically, the backstepping kernels used to construct the control law  are approximated with neural operators (NO) in order to improve computational efficiency. The key challenge lies in deriving the stability conditions with respect to the Markov-jumping parameter uncertainty and NO approximation errors. The mean-square exponential stability of the stochastic system is achieved through Lyapunov analysis, indicating that the system can be stabilized if the random parameters are sufficiently close to the nominal parameters on average, and NO approximation errors are small enough. The theoretical results are  applied to freeway traffic control under stochastic upstream demands and then validated through numerical simulations.
\end{abstract}

\end{frontmatter}

\section{Introduction}
Boundary control of hyperbolic PDEs is widely applied to engineering  problems that require point actuation for spatial-temporal stabilization, such as oil drilling~\cite{wang2020delay}, traffic flow~\cite{yu2022traffic}, gas pipes~\cite{bastin2016stability}. Lyapunov-based control methods are widely applied including PI control~\cite{zhang2019pi}, feedback control~\cite{karafyllis2018feedback} and  backstepping approach~\cite{krstic2008boundary}.  The PDE backstepping achieves Lyapunov stabilization by Volterra spatial transformation and then eliminates destabilizing in-domain terms by boundary feedback controller design. It involves solving kernel equations for the invertible backstepping transformation, which can be time-consuming and difficult for practical implementation. Over recent years, machine learning (ML) methods such as Physics-informed Neural Networks (PINN)~\cite{karniadakis2021physics} and Reinforcement Learning (RL)~\cite{yu2021reinforcement} have been applied to  develop efficient learning-based models to solve PDEs and to accelerate the computation speed. But they suffer from the generalization issues for change of model parameters and initial conditions. Neural operators (NO) have been proposed to learn the operator mappings of functionals~\cite{lu2021learning} and then were applied to obtain stability-guaranteed backstepping controllers~\cite{bhan2023neural}. In this paper, we will study operator learning for stabilization of Markov-jumping hyperbolic PDEs,  and the robustness of NO-controller to stochastic parameters. 

We consider hyperbolic PDE systems with stochastic parameters that are governed by a Markov chain. Stability analysis and control problem of Markov-jumping hyperbolic PDEs have been widely investigated~\cite{amin2011exponential,bolzern2006almost,wang2012stochastically,zhang2017stochastic,prieur2014stability}. The parameters uncertainty is initially represented by switching signals that are defined as a piecewise constant function and right-continuous. 
The author dealt with the stochastic delays and then converted the delayed system into a PDE-ODE system and designed the controller to robustly compensate for the stochastic delay using the backstepping method~\cite{kong2022prediction}. 
For application in traffic flow control, Zhang et al.~\cite{zhang2017stochastic} designed a boundary feedback law to stochastically exponentially stabilize the traffic flow whose dynamics are governed by conservation laws. Our previous work investigated the mean-square exponential stability of the mixed-autonomy traffic system with Markov-jumping parameters, and the controller was designed by the backstepping method~\cite{zhang2023mean}. However, both the Lyapunov design and the backstepping method for PDEs with parameter uncertainties suffer the high computational cost problem as solving LMIs and backstepping kernels is time-consuming. Therefore, it is relevant to adopt machine learning tools to accelerate the computations for controllers that are robust to the parameter uncertainties. 

Backstepping control for PDEs was first proposed in~\cite{krstic2008boundary} with full-stated feedback control laws and output-feedback control laws. Vazquez et al.~\cite{vazquez2011backstepping} proposed a boundary controller for a $2\times 2$ hyperbolic system and the well-posedness of the kernel equations used for backstepping transformation was proved. Stabilization of higher-order hyperbolic PDE systems, such as the $n + 1$ system and the $n + m$ system, was solved in~\cite{hu2015control,di2013stabilization}. 
The reader is referred to~\cite{vazquez2024backstepping} for a survey on backstepping. Recently, the mean-square exponential stabilization of coupled hyperbolic systems with random parameters was addressed in~\cite{auriol2023mean}. More precisely, it was shown through a Lyapunov analysis, that a nominal backstepping controller was robust to random system parameter perturbations, provided the nominal parameters are sufficiently close to the stochastic ones on average. 

With applications in traffic congestion problem, Yu et al. first applied the backstepping control method for the Aw–Rascle–Zhang~(ARZ) traffic model of the hyperbolic PDE type~\cite{aw2000resurrection,zhang2002non} and then extended the result to two-class traffic, two-lane traffic, and cascaded traffic control~\cite{yu2022traffic}. However, the aforementioned backstepping design for hyperbolic PDEs needs to take a backstepping transformation and solve kernel equations(another PDE) that are induced by the backstepping transformation. Solving kernel equations is time-consuming and requires an intensive depth of expertise in the PDE field. Although an explicit solution can be obtained using the power series~\cite{vazquez2014marcum,vazquez2023power}, it needs to define appropriate power series and prove their convergence for the exact kernels. It may cause higher computational burden.

Recently, an increasing number of learning-based methods are applied to solve PDEs and control problems. PINN is proposed for learning the dynamics of PDEs and solving the forward and inverse problems of nonlinear PDEs~\cite{karniadakis2021physics}. But PINNs need retraining when initial conditions change, which also brings the problem of increased training time and complexity of training settings. It only works in a specific set of parameters. While RL lacks a theoretical guarantee of exponential stability. The adaptability of PINN and RL can be poor for control of PDEs under different conditions, in particular, the problem under stochastic system parameters considered in this paper. 

Compared with PINN and RL, NO exhibits the ability to learn operator mapping of functionals, which makes it quite efficient to solve the boundary control problem of PDEs~\cite{lu2021learning}. Especially in approximating the backstepping kernels, the exponential stability of the closed loop is guaranteed through the theoretical derivation. Bhan et al.~\cite{bhan2023neural, bhan2023operator} adopted NO to accelerate computation speed for obtaining control gains and control laws. For hyperbolic PDEs, Wang et al.~\cite{wang_backstepping_2025} adopted neural operators to approximate the backstepping kernels and provided the stability of the 2 $\times$ 2 hyperbolic PDEs under neural operators. All the previous results of NO above focused on adopting NO for control of the deterministic PDEs. The stability results of Markov-jumping hyperbolic PDEs under NO have not been explored. In this paper, we investigated the robust stabilization of NO for the Markov-jumping hyperbolic PDEs. It was adopted to approximate the backstepping kernels and then the stability of the Markov-jumping hyperbolic PDEs with NO-approximated kernels was analyzed.

The main contributions of this paper are twofold. First, we propose a NO-approximated controller that guarantees robust stabilization for hyperbolic PDEs with Markov-jumping parameters. To the best of our knowledge, this is the first theoretical result establishing the use of operator learning for the robust control of linear Markov-jumping hyperbolic PDEs. {Different with the exponential stability result in~\cite{amin2011exponential} was obtained through the condition that the spectral radius of boundary coupling matrices should satisfy specific conditions for all switch modes, this paper does not add constrains on the boundary coupling coefficients thus we get the mean-square exponential stability of the PDE system.} Second, we demonstrate the applicability of our approach through a traffic congestion control problem, addressing freeway regulation under stochastic upstream demands. The use of neural operators not only improves the computational efficiency of solving PDEs but also ensures both system stability and solution accuracy. Methodologically, the paper extends the Lyapunov analysis proposed in~\cite{auriol2023mean} to encompass NO-approximations.


\textbf{Notation}: We denote $L^2([0,1],\mathbb{R})$ the space of real-valued square-integrable functions defined on $[0,1]$ with standard $L^2$ norm, i.e., for any $f \in L^2([0,1],\mathbb{R})$, we have $\norm{f}_{L^2} = \left( \int_0^1 f^2(x)dx \right)^{\frac{1}{2}}$. The supremum norm is $\norm{\cdot}_{\infty}$.{$\norm{\cdot}$ denotes the standard Euclidean norm}. $\mathbb{E}(x)$ denotes the expectation of a random variable $x$. For a random signal $x(t)$, we denote the conditional expectation of $x(t)$ at the instant $t$ with initial condition $x(0)$ at instant $s \leq t$ as $\mathbb{E}_{[s,x(0)]}(x(t))$.  {The set $\mathcal{C}^n([0,1]), n\in \mathbb{N}$ denotes the space of real-valued functions defined on $[0,1]$ that are $n$ times differentiable and whose $n^{\text{th}}$ derivative is continuous.}

\section{Problem statement}\label{sec2}
\subsection{System with Markov-jumping parameter uncertainties}
We consider a stochastic $2 \times 2$ linear hyperbolic system
\begin{align}
 \partial_t u(x, t)+\lambda(t) \partial_x u(x, t)&=\sigma^{+}(t) v(x, t),\label{stosys1}\\
 \partial_t v(x, t)-\mu(t) \partial_x v(x, t)&=\sigma^{-}(t) u(x, t),
\end{align}
with boundary conditions 
\begin{align}
 u(0,t)&=\varphi(t) v(0, t), \\
 v(1,t)&=\varrho(t) u(1, t)+U(t), \label{stosys4}
\end{align}
where the spatial and time variables $(x,t)$ belong to $\{[0,1]\times \mathbb{R}^+ \}$. 
The stochastic characteristic speeds $\lambda(t)>0$ and $\mu(t)>0$ are time-varying. The in-domain couplings $\sigma^{+}(t)$, $\sigma^{-}(t)$ and boundary couplings $\varphi(t)$, $\varrho(t)$ are also stochastic and time-varying. The different parameters are random independent variables. The set of the random variables is denoted as $\mathfrak{S} = \{\lambda, \mu, \sigma^+, \sigma^-, \varphi,\varrho\}$. Each random element $X$ of the set $\mathfrak{S}$ is a Markov process with the following properties:
\begin{enumerate}
    \item $X(t) \in\left\{X_i, i \in\left\{1, \ldots, r_X\right\}\right\}$, $r_X \in \mathbb{N}$ with $\underline{X} \leq X_1<\cdots<X_{r_X} \leq \bar{X}$.
    \item The transition probabilities $P_{ij}^X(t_1,t_2)$ describes the probability to switch from $X_i$ at time $t_1$ to $X_j$ at time $t_2$. The $i,j$ are also in the finite modes of the Markov process with $((i, j) \in\left\{1, \ldots, r_X\right\}^2, 0 \leq t_1 \leq t_2)$. In addition, $P_{ij}^X(t_1,t_2)$ satisfies $P_{i j}^X: \mathbb{R}^2 \rightarrow[0,1]$ with $\sum_{j=1}^{r_X} P_{i j}^X\left(t_1, t_2\right)=1$. $P_{i j}^X$ follows the Kolmogorov equation~\cite{kolmanovsky2001mean,hoyland2009system,ross2014introduction}
        \begin{align}\label{kolman}
            & \partial_t P_{i j}^X(s, t)=-c_j^X(t) P_{i j}^X(s, t)+\sum_{k=1}^{r_X} P_{i k}^X(s, t) \tau_{k j}^X(t) \nonumber \\
            & P_{i i}^X(s, s)=1, \text { and } P_{i j}^X(s, s)=0 \text { for } i \neq j, 
        \end{align}
    where $\tau_{ij}$ and $c_j^X =\sum_{k=1}^{r_X} \tau_{jk}^X$ are non-negative valued functions such that for any $t$, $\tau_{ii}^X=0$. The functions $\tau_{ik}^X$ are bounded by a constant $\tau_{X}^\star$.
    \item The realizations of $X$ are right-continuous.
\end{enumerate}

We assume that the lower bounds for the characteristic speed $\underline{\lambda}$, $\underline{\mu}$ are positive. For each $X \in \mathfrak{S}$, we define $T_X \in \{X_1,\ldots, X_{r_X} \}$ as the set of realizations for the variable $X$.  {We denote by $r_X:=|T_X|$ the number of modes of $X\in \mathfrak{S}$. Since the parameters are independent, the joint mode set is the Cartesian product $\mathfrak{R}:=\{1,\dots,r_\lambda\}\times\{1,\dots,r_\mu\}\times\{1,\dots,r_{\sigma^+}\}\times\{1,\dots,r_{\sigma^-}\}\times\{1,\dots,r_\phi\}\times\{1,\dots,r_\varrho\}$, whose cardinality is $r = r_\lambda \times r_\mu \times r_{\sigma^+} \times r_{\sigma^-} \times r_\varphi \times r_\varrho$.} 
Let $\delta(t)\in (\mathbb{R}^+)^2 \times \mathbb{R}^4$ be defined by
\begin{align}
    \delta(t)=\left(\lambda(t), \mu(t), \sigma^{+}(t), \sigma^{-}(t), \varphi(t), \varrho(t)\right).
\end{align}
 {$\delta(t)$ is a set including all Markov-jumping parameters and it is also a big Markov process due to the independence of different system parameters. The transition probabilities are obtained from the those of $\mathfrak{S}$.}
We also define the modes indices of each element in $\mathfrak{S}$ as $j \in \{1,\ldots, r_X\}, X \in \mathfrak{S}$, i.e., $\delta(t) = \delta_j$ means the elements $X$ is mode $j$ at time instant $t$, $X(t) = X_{j_X}$.

\subsection{NO-approximated nominal control law}
Let us first consider the following 2 $\times$ 2 linear hyperbolic system without stochastic uncertainty, called nominal system
\begin{align}
    \partial_t u_{nom}(x,t) + \lambda_0 \partial_x u_{nom}(x,t) &= \sigma^+_0 v_{nom}(x,t), \label{nom1}\\
    \partial_t v_{nom}(x,t) - \mu_0 \partial_x v_{nom}(x,t) &= \sigma^-_0 u_{nom}(x,t),
\end{align}
with boundary conditions
\begin{align}
    u_{nom}(0,t) &= \varphi_0 v_{nom}(0,t),\\
    v_{nom}(1,t) &= \varrho_0 u_{nom}(1,t) + U(t), \label{nom4}
\end{align}
where the nominal characteristic speeds $\lambda_0 > 0$ and $\mu_0 > 0$ are constant. In-domain couplings $\sigma^+_0$, $\sigma^-_0$ and boundary couplings $\varphi_0$, $\varrho_0$ are also assumed to be constant. We also define the nominal set  {$\delta_0 = (\lambda_0, \mu_0,\sigma^+_0, \sigma^-_0, \varphi_0,\varrho_0) \in (\mathbb{R}^+)^2\times \mathbb{R}^4$} including all the nominal parameters.  {Let $\mathcal{U}\subset ( \mathbb{R}^+ )^2\times\mathbb{R}^4$ denote a bounded, closed set of admissible nominal parameters on which we carry out both analysis and learning as $\mathcal{U}:=\{\delta_0 = (\lambda_0, \mu_0,\sigma^+_0, \sigma^-_0, \varphi_0,\varrho_0): \lambda_0\in [\underline{\lambda},\overline{\lambda}], \mu_0 \in [\underline{\mu},\overline{\mu}], \abs{\sigma_0^\pm}\leq\overline{\sigma},\abs{\varphi_0}\leq\overline{\varphi},\abs{\varrho_0}\leq \overline{\varrho}\}$with fixed constants $0<\underline{\lambda}\leq\overline{\lambda}$, $0<\underline{\mu}\leq\overline{\mu}$ and positive bounds $\overline{\sigma},\overline{\phi},\overline{\varrho}$. All regularity and Lipschitz constants below depend only on $\mathcal{U}$.}  
It corresponds to the system parameters under the nominal mode. The function $U(t)$ is the boundary control input that is given as:
\begin{align}\label{nom_control}
    U(t)  =& - \varrho_0 u_{nom}(1,t) + \int_0^1 K^{v u}(1, \xi) u_{n o m}(\xi,t) d \xi \nonumber \\
    & +\int_0^1 K^{v v}(1, \xi) v_{n o m}(\xi,t) d \xi.
\end{align}
where the backstepping kernels $K^{vu},K^{vv}$$\in \mathcal{C}^1$  are defined  on the triangular domain $\mathcal{T} = \{ 0 \leq \xi \leq x \leq 1\}$. 
The kernels can be obtained by solving kernel equations in~\cite{vazquez2011backstepping}. 
Using the control law~\eqref{nom_control}, the closed-loop system \eqref{nom1}-\eqref{nom4} is well-posed and exponentially stable in $L^2$ norm~\cite{bastin2016stability,coron2021boundary}. We have the following theorem.
\begin{thm}\cite[Theorem 1]{vazquez2011backstepping}
    Consider the system~\eqref{nom1}-\eqref{nom4} with initial conditions $u_{nom}^0$ and $v_{nom}^0$, and control law~\eqref{nom_control}, then the equilibrium $u\equiv v \equiv 0$ is exponentially stable in the $L_2$ sense, the equilibrium is reached at finite time $t_f = \frac{1}{\lambda_0} + \frac{1}{\mu_0}$.
\end{thm}
We now introduce an operator that maps (for a given nominal set of parameters $\delta_0$) the nominal system parameters to the corresponding backstepping kernels. This leads to the following lemma:
 {
\begin{lem}\label{defkernel}
    The kernel operator $\mathcal{K}$: $\mathcal{U}$ $\to$ $(\mathcal{C}^1(\mathcal{T}))^4$ defined by $\mathcal{K}(\delta_0)(x,\xi) =:(K^{uu},K^{uv},K^{vu},K^{vv})$ is locally Lipschitz. More precisely, there exists a constant $c_{\mathcal{U}}>0$, depending only on the set $\mathcal{U}$, such that for all $\delta_a$ and $\delta_b \in \mathcal{U}$, we have
    \begin{align}\label{kernelop}
        \norm{\mathcal K(\delta_a)-\mathcal K(\delta_b)}_{(\mathcal{C}^1(\mathcal T))^4} \leq c_\mathcal{U} \norm{\delta_a-\delta_b},
    \end{align}
    where $\norm{\cdot}_{\mathcal{C}^1(\mathcal{T})}:= \norm{\cdot}_{L^\infty(\mathcal{T})} + \norm{\partial_x(\cdot)}_{L^\infty(\mathcal{T})} + \norm{\partial_\xi(\cdot)}_{L^\infty(\mathcal{T})}$.
\end{lem}
\begin{pf}
    For each fixed $\delta_0$, the coupled kernel equations associated with~\eqref{kernelop}  admit a unique solution $\mathcal K(\delta_0)\in\big(\mathcal{C}^1(\mathcal T)\big)^4$ and depend $\mathcal{C}^1$-smoothly on the parameters~\cite{vazquez2011backstepping,wang_backstepping_2025,qi2024neural}. Moreover, on any bounded parameter set $\mathcal{U} \subset (\mathbb R^+)^2\times\mathbb R^4 $ there exists $M_\mathcal{U}>0$ such that $\norm{\mathcal{K}(\delta_0)}_{(\mathcal{C}^1)^4} \leq M_\mathcal{U}$, for all $\delta_0 \in 
    \mathcal{U}$~\cite{di2013stabilization,wang_backstepping_2025}. 
    Denote $K_a^{\cdot\cdot}$ (resp. $K_b^{\cdot\cdot}$) the kernels  associated with a set of parameters $\delta_a$ (resp. $\delta_b$). Denote $\Delta K^{\cdot\cdot}:=K^{\cdot\cdot}_a-K^{\cdot\cdot}_b$.
    Integrating along characteristic lines yields  and taking $\mathcal{C}^1(\mathcal{T})$-norms 
    gives the following estimate~\cite{vazquez2011backstepping}:
    \begin{align*}
        \norm{\Delta K^{\cdot\cdot}}_{\mathcal{C}^1(\mathcal{T})} \leq c_2^\mathcal{U}\norm{\delta_a - \delta_b} + c_3^\mathcal{U} \int_\xi^x \norm{\Delta K^{\cdot\cdot}}_{\mathcal{C}^1(\mathcal{T}_s)} ds,
    \end{align*}
    where $c_2^\mathcal{U}, c_3^\mathcal{U} > 0$ and $\mathcal{T}_s := \{ (\mathfrak{y},\mathfrak{z})\in\mathcal{T}:\mathfrak{z} \leq \mathfrak{y} \leq s \}$. The Volterra–Gr\"{o}nwall inequality on $\mathcal T$ then yields
    \begin{align}
        \norm{\Delta K^{\cdot\cdot}}_{\mathcal{C}^1(\mathcal{T})} \leq c_\mathcal{U} \norm{\delta_a - \delta_b},
    \end{align}
    for some $c_\mathcal{U} > 0$. Applying this to each of the four components, we have $\norm{\mathcal K(\delta_a)-\mathcal K(\delta_b)}_{(\mathcal{C}^1(\mathcal T))^4} \leq c_\mathcal{U} \norm{\delta_a-\delta_b}$. This completes the proof of Lemma~\ref{defkernel}. 
\end{pf}
\begin{rem}
    The local Lipschitz continuity of $\mathcal K$ established above implies that, on any bounded parameter set $\mathcal{U}$, $\mathcal K$ is continuous as a map into $(\mathcal{C}^1(\mathcal T))^4$.  Consequently, universal approximation results for operator-learning architectures (e.g., DeepONet) guarantee the existence of a neural operator to approximate the mapping.
\end{rem}
}
From Lemma~\ref{defkernel}, the kernel operator $\mathcal{K}$ maps the system parameters to the backstepping kernels, such that there exists a neural operator approximating the kernel operator $\mathcal{K}$, then we have the following lemma:
\begin{lem} \label{kerapplem}
    For all $\epsilon > 0$, there exists a neural operator $\hat{\mathcal{K}}$ such that for all $(x,\xi) \in \mathcal{T}$,
        \begin{align}\label{neuraloperator}
    \sup_{\delta_0 \in \mathcal{U}}\norm{ \mathcal{K}(\delta_0)(x,\xi)  - \mathcal{\Hat{K}}(\delta_0)(x,\xi)} < \epsilon.
    \end{align}
\end{lem}
\begin{pf}
    The proof could be easily obtained using the results in Lemma~\ref{defkernel} and following same steps in~\cite{bhan2023neural,deng2022approximation}.
\end{pf}
\begin{rem}
    The maximum approximation error $\epsilon$ is defined as the error between the NO-approximated kernels and the exact kernels. The error is related to the network size, neural layers and neurons in each layer of the designed network. The selection of these parameters is empirical. Theoretically, the $\epsilon$ can be chosen small enough given enough computing resources.
\end{rem}
Using the neural operator $\hat{\mathcal{K}}$, we can easily obtain the NO-approximated nominal control law 
\begin{align}\label{NO_nom_control}
    U_{NO}(t)  =& - \varrho_0 u_{nom}(1,t) + \int_0^1 \hat{K}^{v u}(1, \xi) u_{n o m}(\xi,t) d \xi \nonumber \\
    & +\int_0^1 \hat{K}^{v v}(1, \xi) v_{n o m}( \xi,t) d \xi.
\end{align}
Now we first state the well-posedness of the closed-loop system with Markov-jumping parameters under the NO-approximated kernels. We have the following Lemma:
\begin{lem}\label{well-posedness}
    For any initial conditions of the stochastic system $u^0(x),v^0(x)\in L^2([0,1],\mathbb{R}^2)$ and any initial states $\delta(0)$ for the stochastic parameters, the system~\eqref{stosys1}-\eqref{stosys4} with the NO-approximated nominal control law~\eqref{NO_nom_control} has a unique solution such that for any $t$,
    \begin{align}\label{exptation}
        \mathbb{E}_{[0,(u^0(x),v^0(x),\delta(0))]}(\norm{u(\cdot,t),v(\cdot,t)}_{L^2}^2) < \infty.
    \end{align}
\end{lem}
\begin{pf}
    The proof of the well-posedness of the system under the NO-approximated backstepping kernels can be easily obtained by extending the results from~\cite{auriol2023mean,zhang2023mean,zhang2017stochastic}. For every event of the stochastic process $X(t)$, $t\leq 0$ is a right-continuous function with a finite number of jumps in a finite time interval. So there exists a sequence $\{t_k: k = 0,1,\ldots\}$ such that $t_0=0$, $\lim_{t\to \infty}, t_k \to \infty$. Starting from the initial time instant, we can fix the random parameter at the first time interval. The control law is obtained by~\eqref{NO_nom_control}. 
    We have stated the well-posedness and the regularity of the NO-approximated backstepping kernels in Lemma~\ref{kerapplem}. The NO-approximated backstepping kernels have the same functional form with the nominal kernels. Therefore, the initial-value problem of system~\eqref{stosys1}-\eqref{stosys4} under the control law~\eqref{NO_nom_control} has one, and only one solution using the results in~\cite[Theorem A.4]{bastin2016stability} and \cite[Appendix. A]{coron2021boundary} as the system in this paper is a particular case of them. Then iterating the process for each time interval on the whole time domain, we can get the stochastic system has a unique solution for any $t\geq 0$ that satisfies~\eqref{exptation}.
    This completes the proof of Lemma~\ref{well-posedness}. 
\end{pf}
\subsection{Main results}
In this section, we state the main results of our paper. The objective is to prove that the NO-approximated nominal control law~\eqref{NO_nom_control} can still stabilize the stochastic system~\eqref{stosys1}-\eqref{stosys4}, providing the nominal parameters are sufficiently close to the stochastic ones on average and a small approximation error $\epsilon$. 
In other words, we want to show the following robust stabilization result
\begin{thm}\label{mainthm}
    There exists a constant $\phi^\star>0$ and a small enough approximation error $\epsilon>0$, if for all time $t \geq 0$ and $X \in \mathfrak{S}$,
    \begin{align}
        \sum_{X \in \mathfrak{S}} \mathbb{E}_{[0, X(0)]}\left(\abs{X^0-X(t)}\right) \leq \phi^\star,
    \end{align}
    the closed-loop system~\eqref{stosys1}-\eqref{stosys4} with the control law~\eqref{NO_nom_control} is mean-square exponentially stable, namely, there exist constants $\kappa = \kappa(\phi^\star)>0$, $\varsigma=\varsigma(\epsilon)>0$, independent of $t$, such that
    \begin{align}
        \mathbb{E}_{[0,(w(x, 0),\delta(0))]}(\norm{w(x,t)}_{L^2}^2) \leq \kappa(\phi^\star) \mathrm{e}^{-\varsigma(\epsilon) t} \norm{w(x,0)}_{L^2}^2,
    \end{align}
    where $w(x,t) = (u(x,t), v(x,t)) \in (L^2([0,1],\mathbb{R}))^2$.
\end{thm}
\begin{rem}
   Compared to~\cite{auriol2023mean}, the system considered in this work involves two sources of uncertainty: the Markov-jumping parameters and the approximation error introduced by the neural operator. These two uncertainties, denoted respectively by $\phi$ and $\epsilon$, are  independent. In the proof of the main theorem, explicit bounds are provided for both the Markov-jumping variation, denoted $\phi^\star$, and the approximation error $\epsilon$. Due to the inherent conservatism of the Lyapunov-based analysis, the bound $\phi^\star$ is mainly of practical relevance. The stated bounds are conservative, and the result should be interpreted qualitatively, establishing the existence of robustness margins. In particular, a smaller value of $\phi^\star$ leads to faster convergence of the stochastic system, and the same holds for the approximation error: reducing $\epsilon$ improves the convergence rate.
\end{rem}

\section{NO-approximated kernels for stochastic system}\label{sec3}
In this section, we give the details of the backstepping transformation and derive the target system under NO-approximated nominal control law.
Following the backstepping method proposed in~\cite{vazquez2011backstepping}, 
\begin{align}
& \alpha(x, t)=u - \int_0^x K^{u u}(x, \xi) u+K^{u v}(x, \xi) v\, d \xi, \label{backtrans1}\\
& \beta(x, t)=v- \int_0^x K^{v u}(x, \xi) u+K^{v v}(x, \xi) v\, d \xi,\label{backtrans2}
\end{align}
where these kernels $K^{uu},K^{uv} \in \mathcal{C}^1$ defined on the same triangular domain $\mathcal{T}$ are obtained by solving the associated kernels equations in~\cite{vazquez2011backstepping}. Then
we get the following stochastic target system,
\begin{align}
    &\partial_t \alpha (x,t) + \lambda(t) \partial_x \alpha(x,t) = f_1(\delta(t)) v(x,t) + f_2(\delta(t)) \beta(0,t) \nonumber\\
    &+ \int_0^x f_3(\delta(t),x,\xi)u(\xi,t) +f_4(\delta(t),x,\xi)v(\xi,t) d \xi, \label{tarsto1}\\
    &\partial_t \beta(x,t)-\mu(t) \partial_x \beta(x,t)=g_1(\delta(t)) u( x,t)+g_2(\delta(t)) \beta(0,t)\nonumber\\
    &+\int_0^x g_3(\delta(t), x, \xi) u(\xi, t)+g_4(\delta(t), x, \xi) v(\xi,t) d \xi,
\end{align}
with boundary conditions
\begin{align}
    &\alpha(0,t) = \varphi(t) \beta(0,t),\\
    &\beta(1,t) = (\varrho(t)-\varrho_0) u(1,t) \nonumber\\
    &-\int_0^1 \left( {K}^{v u}(1, \xi) - \Hat{K}^{v u}(1, \xi) \right)  u(\xi, t)\nonumber\\
    &+\left( {K}^{v v}(1, \xi) - \Hat{K}^{v v}(1, \xi)  \right)v(\xi, t) d \xi,\label{tarsto4}
\end{align}
where $f_1 =\left( \sigma^{+}(t)-\sigma_0^{+} \frac{\lambda(t)+\mu(t)}{\lambda_0+\mu_0} \right)$, $f_2 = \left(\mu(t) - \frac{\lambda(t) \varphi(t)\mu_0}{\lambda_0 \varphi_0}\right)$ $K^{u v}(x, 0)$, $f_3 = \left( \frac{\lambda(t)}{\lambda_0} \sigma_0^{-}-\sigma^{-}(t) \right)K^{u v}(x, \xi)$, $f_4 =\left(\lambda_0-\lambda(t)\right) \partial_x K^{u v}(x, \xi)+\left(\mu(t)-\mu_0\right) \partial_\xi K^{u v}(x, \xi)- \left(\sigma^{+}(t)-\sigma_0^{+}\right)K^{u u}(x, \xi)$, and $g_1=\sigma^-(t) - \frac{\lambda(t)+\mu(t)}{\lambda_0+\mu_0}\sigma^-_0$, $g_2=\left( -\lambda(t)\varphi(t) + \mu(t)\frac{\lambda_0\varphi_0}{\mu_0} \right)K^{vu}(x,0)$, $g_3 =(\mu(t) - \mu_0)K^{vu}_x(x,\xi) - (\lambda(t) - \lambda_0)K^{vu}_\xi(x,\xi)-(\sigma^-(t) - \sigma^-_0)K^{vv}(x,\xi)$, $g_4 =\left( \frac{\sigma^+_0 \mu(t)}{\mu_0} - \sigma^+(t) \right)K^{vu}(x,\xi)$.
The backstepping transformation~\eqref{backtrans1}-\eqref{backtrans2} is a Volterra type so that it is boundedly invertible~\cite{yoshida1960lectures}. Therefore, all the terms that depend on $(u,v)$ in the target system could be expressed in terms of $(\alpha,\beta)$ using the inverse backstepping transformation associated with~\eqref{backtrans1}-\eqref{backtrans2}. We chose to keep them as functions of $(u,v)$ to avoid complex expressions. As it will be seen, these expressions are convenient for the robust analysis. Thus,
the target system is simpler in the sense that it
simplifies the robustness analysis that will be carried out in the next section. 
Next, we bound all the terms in the target stochastic system.  Due to the invertibility of the backstepping transformation~\eqref{backtrans1}-\eqref{backtrans2}, the states of the stochastic target system and the original states have equivalent $L^2$ norms, namely, there exist two constants $m_1>0$ and $m_2>0$ such that $m_1\norm{w(x,t)}^2_{L^2} \leq \norm{\Xi(x,t)}^2_{L^2} \leq m_2 \norm{w(x,t)}^2_{L^2}$, where $\Xi(x,t) = (\alpha(x,t), \beta(x,t)) \in (L^2([0,1],\mathbb{R}))^2$. 
We have the following lemma:
\begin{lem}\label{boundf}
    There exists a constant $M_0 > 0$, such that for any realization of $X(t) \in \mathfrak{S}$, for any $(x,\xi) \in \mathcal{T}$, and $i\in\{1,2,3,4\}$, we have the following bound
    \begin{align}
        \abs{f_{i}(\delta(t))} & \leq M_0 \sum_{X\in \mathfrak{S}} \abs{X^0 - X(t)}, \\ 
        \abs{g_i(\delta(t))} & \leq M_0 \sum_{X\in \mathfrak{S}} \abs{X^0 - X(t)}.
    \end{align}
\end{lem}
\begin{pf}
    For the bound of the functions, we have $\abs{f_1(\delta(t))} \leq \max\{1, \abs{\frac{\sigma^+_0}{\lambda_0+\mu_0}}\} \sum_{X\in \mathfrak{S}} \abs{X^0 - X(t)}$. For the function $f_2(\delta(t))$, $\abs{f_2(\delta(t))} \leq  \max\{ 1,\frac{\mu_0}{\lambda_0},\frac{\mu_0\overline{\lambda}}{\lambda_0\varphi_0} \}\\ \sup_{\mathcal{T}} \norm{K^{uv}(x,0)} \sum_{X\in \mathfrak{S}} \abs{X^0 - X(t)}$. For function $f_3(\delta(t))$, we have $ \abs{f_3(\delta(t))} \leq  ( 1+\frac{\abs{\sigma^-_0}}{\lambda_0})\sup_{\mathcal{T}}\norm{K^{uv}(x,\xi)}\\\sum_{X\in \mathfrak{S}} \abs{X^0 - X(t)}$. Using the same method, for $f_4(\delta(t))$, we get $\abs{f_4(\delta(t))} \leq \max\{ \sup_{\mathcal{T}}\norm{\partial_x K^{uv}(x,\cdot)},\\\sup_{\mathcal{T}}\norm{\partial_{\xi} K^{uv}(\cdot,\xi)}, \sup_{\mathcal{T}}\norm{K^{uv}(\cdot,\cdot)}\}\sum_{X\in \mathfrak{S}} \abs{X^0 - X(t)}$. The backstepping kernels $K^{\cdot \cdot}$ are well-defined and bounded such that their derivatives are also bounded and well-defined. For the functions of $g_i(\delta(t))$, we can use the same method to derive the bound of them. This finishes the proof of Lemma~\ref{boundf}.
\end{pf}

\section{Lyapunov analysis for the stochastic system}\label{sec4}
In this section, we will provide the Lyapunov analysis of the system under the NO-approximated control law to show that the system is mean-square exponentially stable. The objective is to prove Theorem~\ref{mainthm}. To do that, we first need to define the Lyapunov candidate and then conduct the Lyapunov analysis to finish the proof.

\subsection{Derivation of Lyapunov functional}
The previous section has proved that the nominal system with nominal controller is exponentially stable. For the stochastic target system, we consider the following stochastic Lyapunov functional candidate  as
\begin{align}\label{lyapunv}
    V(\Xi,\delta) = \int_0^1 \frac{\mathrm{e}^{-\frac{\nu}{\lambda(t)}}}{\lambda(t)} \alpha^2(x,t) + a \frac{\mathrm{e}^{\frac{\nu}{\mu(t)}}}{\mu(t)} \beta^2(x,t) dx.
\end{align}
If the system stays at mode $j$ where $\delta(t) = \delta_j$, the Lyapunov candidate can be also written in $V_j(t) =  \int_0^1 \frac{\mathrm{e}^{-\frac{\nu}{\lambda_j}}}{\lambda_j} \alpha^2(x,t) + a \frac{\mathrm{e}^{\frac{\nu}{\mu_j}}}{\mu_j} \beta^2(x,t) dx$. The Lyapunov candidate is equivalent to the $L^2$ norm of state $\Xi(x,t)$, there exist two constants $m_3>0$ and $m_4>0$ that $m_3 \norm{\Xi}^2_{L^2} \leq V(\Xi,\delta(t)) \leq  m_4 \norm{\Xi}^2_{L^2}$.
And then we consider the infinitesimal generator $L$ of the Lyapunov candidate $V$ defined in~\eqref{lyapunv} as~\cite{ross2014introduction}
\begin{align}
 L V(\Xi,\delta) =&\limsup _{\Delta t \rightarrow 0^{+}} \frac{1}{\Delta t} ( \mathbb{E}(V(\Xi(t+\Delta t), \delta(t+\Delta t)))\nonumber\\
&-V(\Xi(t), \delta(t))).
\end{align}
Also, for the infinitesimal generator of the Lyapunov candidate at each Makrov mode $j \in \{1,\dots, r\}$ where $\delta(t) = \delta_j$, we denote 
\begin{align}
L_j V(\Xi)  =\frac{d V}{d \Xi}(\Xi, \delta_j) h_j(\Xi) +\sum_{\ell \in \mathfrak{R}}\left(V_{\ell}(\Xi)-V_j(\Xi)\right) \tau_{j \ell}(t),
\end{align}
where $\ell \in \{1,\dots, r\}$ and the operator $h_j$ is defined as
 {
\begin{align}
    h_j(\Xi) = \begin{bmatrix}
        -\lambda_j \partial_x \alpha(x,t) + f_1(\delta_j) v(x,t) + f_2(\delta_j) \beta(0,t)\\
        +\int_0^x f_3(\delta_j,x,\xi)u(\xi,t) + f_4(\delta_j,x,\xi)v(\xi,t) d \xi\\
         \\
        \mu_j \partial_x \beta(x,t) + g_1(\delta_j)u(x,t) + g_2(\delta_j)\beta(0,t)\\
        +\int_0^x g_3(\delta_j, x, \xi) u(\xi, t)+g_4(\delta_j, x, \xi) v(\xi,t) d \xi
    \end{bmatrix}.
\end{align}
}
To prove the Theorem~\ref{mainthm}, we first give the bound of the probability of the infinitesimal generator of the Lyapunov candidate. We have the following lemma:
\begin{lem}
There exists $\epsilon_0 >0$ such that for all approximation error of NO $0< \epsilon < \epsilon_0$, there exists ${\eta}>0$, $M_2 > 0$, $M_3 >0$ and $m_5 > 0$ such that the Lyapunov functional $V(t)$ satisfies
\begin{align}
&\sum_{j=1}^r P_{i j}(0, t) L_j V(t) \leq -V(t)\Big({\eta}-M_3 \mathcal{Z}(t) \nonumber\\
&-\left(M_2+M_3 r \tau^{\star}\right)\sum_{X\in\mathfrak{S}} \mathbb{E}\left(\abs{X^0 - X(t)}\right)\Big)\nonumber \\
&+ (m_5 \mathbb{E}(\abs{X^0 - X(t)})-\mathrm{e}^{-\frac{\nu}{\bar \lambda}})\alpha^2(1,t),
\end{align}
where the function $\mathcal{Z}(t)$ is defined as:
\begin{align*}
\mathcal{Z}(t)=\sum_{\ell=1}^r\sum_{X\in\mathfrak{S}}\abs{X^0 - X_\ell}\left(\partial_t P_{i \ell}(0, t)+c_\ell P_{i \ell}(0, t)\right).
\end{align*} \label{lem_lyapunov_functional}
\end{lem}
\vspace{-0.5cm}
\begin{pf}
    To prove this lemma, we first compute the term $\frac{d V}{d \Xi}(\Xi, \delta_j) h_j(\Xi)$ of the infinitesimal generator of the Lyapunov candidate. Supposing that the system stays at mode $j$ at some time $t$, such that $\delta(t) = \delta_j$, we have the following result
    \begin{align}\label{derivV}
        &\frac{d V}{d \Xi}(\Xi, \delta_j) h_j(\Xi) = - \nu V_j(t) + \int_0^1 \frac{2}{\lambda_j}\mathrm{e}^{-\frac{\nu}{\lambda_j}x}\alpha(x,t)\nonumber\\
        &\left( f_1(\delta_j) v(x,t) + f_2(\delta_j) \beta(0,t)
        +\int_0^x f_3(\delta_j,x,\xi)u(\xi,t) \right.\nonumber \\
        &+ f_4(\delta_j,x,\xi)v(\xi,t) d \xi \Big)dx + \int_0^1 \frac{2a}{\mu_j}\mathrm{e}^{\frac{\nu}{\mu_j}x}\beta(x,t)\nonumber\\
        &\Big( g_1(\delta_j)u(x,t)  + g_2(\delta_j)\beta(0,t) \nonumber\\
        &\left.+\int_0^x g_3(\delta_j, x, \xi) u(\xi, t)+g_4(\delta_j, x, \xi) v(\xi,t) d \xi \right)dx\nonumber\\
        &+ (\varphi_j^2 - a)\beta^2(0,t) - \mathrm{e}^{-\frac{\nu}{\lambda_j}} \alpha^2(1,t) \nonumber\\
        &+ a \mathrm{e}^{\frac{\nu}{\mu_j}}\Bigg( (\varrho_j - \varrho_0)\alpha(1,t)\nonumber\\
        &+ (\varrho_j - \varrho_0)\int_0^1 K^{uu}(1,\xi)u(\xi,t) + K^{uv}(1,\xi) v(\xi,t) d\xi  \nonumber\\
        &- \int_0^1 ( K^{vu}(1,\xi) - \Hat{K}^{vu}(1,\xi))u(\xi,t) \nonumber\\
        &+(K^{vv}(1,\xi) - \Hat{K}^{vv}(1,\xi))v(\xi,t)d\xi \Bigg)^2.
    \end{align}
    Then we use Young's inequality and the bound of functions in Lemma~\ref{boundf}, for the term $\int_0^1 \frac{2}{\lambda_j}\mathrm{e}^{-\frac{\nu}{\lambda_j}x}\alpha(x,t)$ $f_1(\delta_j) v(x,t) dx$, we have
    \begin{align}
        &\int_0^1 \abs{\frac{2}{\lambda_j}\mathrm{e}^{-\frac{\nu}{\lambda_j}x}\alpha(x,t) f_1(\delta_j) v(x,t)} dx \nonumber \\
        &\leq \frac{1}{\underline{\lambda} m_3}\left(M_0\sum_{X\in \mathfrak{S}} \abs{X^0 - X_j}\right)\left(1+ \frac{1}{m_1}\right)V(t).
    \end{align}
    {We denote $c_i$ arbitrary positive constants in the next.}
    For the second term, using the same method, we get 
    \begin{align}
        &\int_0^1 \abs{\frac{2}{\lambda_j}\mathrm{e}^{-\frac{\nu}{\lambda_j}x}\alpha(x,t) f_2(\delta_j) \beta(0,t)} dx \nonumber\\
        &\leq \frac{1}{\underline{\lambda} m_3 c_1} \left(M_0\sum_{X\in \mathfrak{S}} \abs{X^0 - X_j}\right)V(t) \nonumber\\
        &+ \frac{c_1}{\underline{\lambda}}\left(M_0\sum_{X\in \mathfrak{S}} \abs{X^0 - X_j} \right) \beta^2(0, t).
    \end{align}
    For the third term $f_3(\delta_j)$, we have 
    \begin{align}
        &\int_0^1 \abs{\frac{2}{\lambda_j}\mathrm{e}^{-\frac{\nu}{\lambda_j}x}\alpha(x,t)\int_0^x f_3(\delta_j,x,\xi)u(\xi,t) d\xi } dx \nonumber\\
        & \leq \frac{1}{\underline{\lambda} m_3}\left(M_0\sum_{X\in \mathfrak{S}} \abs{X^0 - X_j}\right) \left(1 + \frac{1}{m_1}\right) V(t).
    \end{align}
    Also, the fourth term $f_4(\delta_j)$ can be bounded by 
    \begin{align}
        &\int_0^1\abs{\frac{2}{\lambda_j}\mathrm{e}^{-\frac{\nu}{\lambda_j}x}\alpha(x,t) \int_0^x f_4(\delta_j,x,\xi) v(\xi,t) d\xi} dx \nonumber\\
        & \leq \frac{1}{\underline{\lambda} m_3} \left(M_0\sum_{X\in \mathfrak{S}} \abs{X^0 - X_j}\right)\left(1 + \frac{1}{m_1}\right) V(t).
    \end{align}
    Next, we will give the bound of the functions $g_1(\delta_j)$ to $g_4(\delta_j)$. Using Young's inequality and the results in Lemma~\ref{boundf} again, we get the results for the four functions. For the first term of $g_1(\delta_j)$, $\int_0^1 |\frac{2a}{\mu_j}\mathrm{e}^{\frac{\nu}{\mu_j}x} \beta(x,t) g_1(\delta_j)u(x,t)| dx\leq \frac{a d_1}{\underline{\mu}m_3}(M_0\sum_{X\in \mathfrak{S}} |X^0 - X_j| )(1 + \frac{1}{m_1}) V(t)$,
    where $d_1$ is the bound of the term $\mathrm{e}^{\frac{\nu}{\mu_j}x}$. 
    For $g_2(\delta_j)$, $\int_0^1 |\frac{2a}{\mu_j}\mathrm{e}^{\frac{\nu}{\mu_j}x} \beta(x,t) g_2(\delta_j)\beta(0,t)| dx \leq \frac{ad_1}{m_3\underline{\mu} c_2} (M_0\sum_{X\in \mathfrak{S}} |X^0 - X_j|)(1 + \frac{1}{m_1}) V(t) + \frac{ad_1c_2}{\underline{\mu}}(M_0 \\\sum_{X\in \mathfrak{S}} |X^0 - X_j|) \beta^2(0,t)$.
    For $g_3(\delta_j)$, $\int_0^1 |\frac{2a}{\mu_j}\mathrm{e}^{\frac{\nu}{\mu_j}x} \beta(x,t)\\\int_0^x g_3(\delta_j, x, \xi) u(\xi, t) d\xi| dx \leq \frac{ad_1}{\underline{\mu}m_3} (M_0\sum_{X\in \mathfrak{S}} |X^0-X_j| )(1 + \frac{1}{m_1}) V(t)$.
    For $g_4(\delta_j)$, $\int_0^1 |\frac{2a}{\mu_j}\mathrm{e}^{\frac{\nu}{\mu_j}x} \beta(x,t)\int_0^x\\ g_4(\delta_j, x, \xi) v(\xi, t) d\xi| dx \leq  \frac{ad_1}{\underline{\mu}m_3} (M_0\sum_{X\in \mathfrak{S}} \abs{X^0 - X_j} )(1 + \frac{1}{m_1}) V(t)$.
    
    Denoting the last term of~\eqref{derivV} as $A=a\mathrm{e}^{\frac{\nu}{\mu_j}}\Big((\varrho_j - \varrho_0)\alpha(1,t)+ (\varrho_j - \varrho_0)\int_0^1 K^{uu}(1,\xi)u(\xi,t)+K^{uv}(1,\xi)$ $v(\xi,t)d\xi$$-\int_0^1( K^{vu}(1,\xi)-\Hat{K}^{vu}(1,\xi))u(\xi,t)+(K^{vv}(1,\xi) - \Hat{K}^{vv}(1,\xi))v(\xi,t)d\xi\Big)^2$, using the basic inequality $(a+b)^2 \leq 2(a^2+b^2)$, and then apply Young's inequality, thus 
    $A \leq 2a \mathrm{e}^{\frac{\nu}{\mu_j}} [((\rho_j - \rho_0)\alpha(1,t)
    + (\rho_j - \rho_0)\int_0^1 K^{uu}(1,\xi)u(\xi,t) + K^{uv}(1,\xi) v(\xi,t) d\xi )^2 + (\int_0^1 ( K^{vu}(1,\xi) - \Hat{K}^{vu}(1,\xi))u(\xi,t) +(K^{vv}(1,\xi) - \Hat{K}^{vv}(1,\xi))v(\xi,t)d\xi )^2]$.
    Expending the first term of $A$ and applying Young's inequality and Cauchy-Schwarz inequality, and then applying the inequality $(a+b)^2 \leq 2(a^2+b^2)$ and maximum approximation error $\epsilon$ to the second term, we get 
    \begin{align*}
        A & \leq 2a \mathrm{e}^{\frac{\nu}{\mu_j}}((1+c_3)(\rho_j - \rho_0)^2 \alpha^2(1,t)+  \frac{2(1+c_3)(\rho_j - \rho_0)^2}{m_1m_3c_3} \nonumber\\
        &\times \max\{\sup_{x\in [0,1]} \norm{K^{uu}(1,x)}^2, \sup_{x\in [0,1]} \norm{K^{uv}(1,x)}^2\} V(t) \nonumber\\
        &+ \frac{2\epsilon^2}{m_1m_3} V(t)).
    \end{align*}
    Therefore, we have the following result,
    \begin{align}
        &\frac{d V}{d \Xi}(\Xi, \delta_j) h_j(\Xi) \leq -\nu V_j(t) + c_4 \epsilon^2 V(t)  \nonumber\\
        & +\left(2 a \mathrm{e}^{\frac{\nu}{\mu_j}}(1+c_3)\bar{\varrho}(\varrho_j - \varrho_0) -  \mathrm{e}^{-\frac{\nu}{\lambda_j}}\right) \alpha^2(1,t)\\
        &+ M_2 \sum_{X\in \mathfrak{S}} \abs{X^0 - X_j} V (t)+ (c_5 + \varphi^2_j - a)\beta^2(0,t)\nonumber,
    \end{align}
    where $M_2  = \frac{3M_0(m_1+1)}{\underline{\lambda}m_1 m_3} + \frac{M_0}{\underline{\lambda}m_3 c_1} + \frac{3 M_0 ad_1 (m_1+1)}{\underline{\mu}m_1 m_3}+ \frac{M_0ad_1(m_1+1)}{\underline{\mu} m_1 m_3 c_2}+ \frac{4a \mathrm{e}^{\frac{\nu}{\mu_j}}\bar{\varrho}(1+c_3)}{m_1m_3 c_3} \times \max\{\sup_{x\in [0,1]} \norm{K^{uu}(1,x)}^2,\\ \sup_{x\in [0,1]} \norm{K^{uv}(1,x)}^2\}, c_4 = \frac{4a \mathrm{e}^{\frac{\nu}{\mu_j}}}{m_1 m_3}, c_5 = (\frac{M_0 c_1}{\underline{\lambda}} + \frac{M_0 ad_1 c_2}{\underline{\mu}})\sum_{X\in\mathfrak{S}}\abs{\overline{X} - \underline{X}}$.
     {There exists a $\eta$ denoting the minimal decay rate of all possible modes for the Lyapunov functionals} such that $-\nu V_j(t) \leq -\eta V(t)$.  {Choosing the designed parameters $c_5$ and $a$, such that $c_5 + \varphi_j^2 -a < 0$ always holds}.
    Then we get the following result
    \begin{align}
        &\frac{d V}{d \Xi}(\Xi, \delta_j) h_j(\Xi) \leq -\bar{\eta} V(t) + M_2 \sum_{X\in \mathfrak{S}} \abs{X^0 - X_j} V (t) \nonumber\\
        & + \left( 2 a \mathrm{e}^{\frac{\nu}{\mu_j}}(1+c_3)\bar{\varrho}(\varrho_j - \varrho_0) -  \mathrm{e}^{-\frac{\nu}{\lambda_j}}\right) \alpha^2(1,t),
    \end{align}
    where $\bar{\eta} = \eta- c_4 \epsilon^2$. The approximation error $\epsilon$ is small enough such that $\bar{\eta}>0 $. 
    For the second term of the infinitesimal generator, applying the mean value theorem to the functions $\lambda \to \frac{\mathrm{e}^{-\frac{\nu}{\lambda}x}}{\lambda}$, $\mu \to \frac{\mathrm{e}^{\frac{\nu}{\mu}x}}{\mu} $, we have $\sum_{\ell \in \mathfrak{R}}\left(V_{\ell}(\Xi)-V_j(\Xi)\right) \tau_{j \ell}(t) \leq M_3 \sum_{\ell = 1}^r\sum_{X\in\mathfrak{S}}\tau_{j\ell}\abs{X_j - X_\ell} V(t)$,
    where $M_3$ is defined by $M_3 = \frac{1}{m_3 \underline{\lambda}^2}\left(\frac{\nu}{\underline{\lambda}} +1 \right) + \frac{1}{m_3 \underline{\mu}^2}\left(\frac{\nu}{\underline{\mu}} + 1\right)\mathrm{e}^{\frac{\nu}{\underline{\mu}}}$.
    So we get the bound of the infinitesimal generator as 
    \begin{align}
         &L_jV(t) \leq -\bar{\eta} V(t) + M_2 \sum_{X\in \mathfrak{S}} \abs{X^0 - X_j} V (t) \nonumber\\
        & +\left( 2 a \mathrm{e}^{\frac{\nu}{\mu_j}}(1+c_3)\bar{\varrho}(\varrho_j - \varrho_0) -  \mathrm{e}^{-\frac{\nu}{\lambda_j}}\right) \alpha^2(1,t)\nonumber\\
        &+ M_3 \sum_{\ell = 1}^r\sum_{X\in\mathfrak{S}}\tau_{j\ell}\abs{X_j - X_\ell} V(t).
    \end{align}
    Next, we will compute the expectation of the infinitesimal generator, defined by $\Bar{L} = \sum_{j=1}^r P_{ij}(0,t)L_jV(t)$.
    Using the triangular inequality and the Kolmogorov equation, we have
    \begin{align}
        \Bar{L} &= \sum_{j=1}^r P_{ij}(0,t)L_jV(t)\nonumber\\
        &\leq -\left(\bar{\eta} - (M_2 + M_3r \tau^\star)\sum_{X\in\mathfrak{S}}\mathbb{E}(\abs{X_j-X^0})\right)V(t)\nonumber\\
        &+ M_3 \sum_{\ell=1}^r\sum_{X\in\mathfrak{S}} \abs{X^0 - X_\ell} (\partial_t P_{ij}(0,t) + c_jP_{ij}(0,t))V(t)\nonumber\\
        &+ \left( m_5 \mathbb{E}(\abs{X^0-X_j}) - \mathrm{e}^{-\frac{\nu}{\lambda_j}}\right)\alpha^2(1,t),
    \end{align}
    where $m_5 = 2a \mathrm{e}^{\frac{\nu}{\underline{\mu}}}(1+c_3)\bar{\varrho}$.
    Let $\mathcal{Z}(t) = \sum_{\ell=1}^r\sum_{X\in\mathfrak{S}}$ $\abs{X^0 - X_\ell} (\partial_t P_{ij}(0,t) + c_jP_{ij}(0,t))$, we get
     {
    \begin{align}
        &\sum_{j=1}^r P_{i j}(0, t) L_j V(t) \leq -V(t)\Big(\bar{\eta}-M_3 \mathcal{Z}(t) \nonumber\\
        &-(M_2+M_3 r \tau^{\star})\sum_{X\in\mathfrak{S}} \mathbb{E}\left(\abs{X^0 - X(t)}\right)\Big)\nonumber \\
        &+ (m_5 \mathbb{E}(\abs{X^0 - X(t)})-\mathrm{e}^{-\frac{\nu}{\bar \lambda}})\alpha^2(1,t).
    \end{align}
    }
    This finishes the proof of Lemma~\ref{lem_lyapunov_functional}.
\end{pf}

\subsection{Mean-square exponential stability}
Previous sections give the bound and expectation of the infinitesimal generator, we will prove the mean-square exponential stability of the stochastic system under the NO-approximated kernels in this section. In this section, we aim to prove the Theorem~\ref{mainthm}. 

Firstly, let us denote $\omega(t) = \bar{\eta}-M_3 \mathcal{Z}(t)-\left(M_2+M_3 r \tau^{\star}\right)$ $\sum_{X\in\mathfrak{S}}\mathbb{E}\left(\abs{X^0 - X(t)}\right)$ and define another functional $\mathcal{H}(t) = \mathrm{e}^{\int_0^t \omega(y)dy} V(t)$. Using Lemma~\ref{lem_lyapunov_functional}, we could always find a $\phi^\star$ smaller than $\frac{\mathrm{e}^{-\frac{\nu}{\bar{\lambda}}}}{m_5}$ , we get the following extended inequality
\begin{align}
    &\sum_{j=1}^r P_{i j}(0, t) L_j V(t) \leq -V(t) \omega(t),
\end{align}
and then we take expectation of the inequality, we find that $\mathbb{E}\left(\sum_{j=1}^r P_{i j}(0, t) L_j V(t)\right) \leq -\mathbb{E}(V(t) \omega(t))$, and then we get that $\mathbb{E}(LV(t)) \leq -\mathbb{E}(V(t)\omega(t))$. So we get $\mathbb{E}(L \mathcal{H}(t)) \leq 0$.
Then we apply the Dynkin's formula~\cite{dynkin2012theory},
\begin{align}
    \mathbb{E}(\mathcal{H}(t)) - \mathcal{H}(0) = \mathbb{E}\left( \int_0^t L\mathcal{H}(y)dy  \right) \leq 0.
\end{align}
For $\mathbb{E}(\mathcal{H}(t))$,
\begin{align}
    \mathbb{E}(\mathcal{H}(t)) \geq \mathbb{E}\left(V(t)\mathrm{e}^{-M_3 \phi^\star + \int_0^t(\bar{\eta} - (M_2 + 2M_3r\tau^\star)\phi^\star)dy}  \right),\label{expectH}
\end{align}
where  {$\phi^\star \geq \frac{\bar{\eta}}{2(M_2 + 2M_3r\tau^\star)}$}, so the following inequality is obtained
\begin{align}
    \mathbb{E}(\mathcal{H}(t)) \geq \mathbb{E}\left( V(t) \mathrm{e}^{-M_3 \phi^\star + \frac{\bar{\eta}}{2}t} \right).
\end{align}
We already know that $ \mathbb{E}(\mathcal{H}(t)) \leq \mathcal{H}(0)$, thus we have
\begin{align}
    \mathbb{E}(V(t)) \leq \mathrm{e}^{M_3 \phi^\star} \mathrm{e}^{-\zeta t} V(0),\label{converge}
\end{align}
where $\zeta = \frac{\bar{\eta}}{2}$. This finishes the proof of the mean-square exponential stability of the stochastic system, namely, the Theorem~\ref{mainthm} is proved because $V(t)$ has the same equivalent norm with $w(x,t)$. $\hfill\blacksquare$
 {
\begin{rem}
    The approximation error $\epsilon$ and the Markov-jumping uncertainty (bounded by $\phi^\star$) originate from different mechanisms: the former arises in the boundary controller due to neural operator approximation, while the latter stems from stochastic mode transitions governed by the Kolmogorov equation. Consequently, their modeling can be decoupled. However, both uncertainties jointly affect the Lyapunov convergence rate, requiring simultaneous smallness for stability. In particular, larger stochastic variations (larger $\phi^\star$) necessitate a smaller $\epsilon$ to preserve the decay rate, indicating an implicit trade-off between the two. The effective convergence rate $\bar{\eta} = \eta - c_4 \epsilon^2$ is a positive constant determined \textit{a priori} by the system parameters and approximation accuracy. It does not depend on the current state and remains constant throughout the system evolution. In the proposed proof, we first tune $\epsilon$ to ensure that $\bar \eta > 0$, and only then bound the stochastic variations to guarantee mean-square stability. Consequently, the maximal admissible bound on the stochastic variations implicitly depends on $\epsilon$. Once $\epsilon$ is chosen, we obtain a quadratic dependency between $\epsilon$ and $\phi^\star$ by $\phi^\star \geq \frac{\bar{\eta}}{2(M_2 + 2M_3r\tau^\star)}$. Alternatively, one could adopt a different strategy by first fixing the maximal bound and then adjusting $\epsilon$ accordingly.
\end{rem}
}

\section{Application to traffic congestion problem}\label{sec5}
\subsection{Traffic flow model}
In this section, we provide the simulation results of the stochastic system with the NO-approximated kernels. 
Considering the following linearized ARZ system
\begin{align}
\partial_t \tilde{q}(x, t)&+v^{\star} \partial_x \tilde{q}(x, t)-\frac{q^{\star}\left(\gamma p^{\star}-v^{\star}\right)}{v^{\star}} \partial_x \tilde{v}(x, t)\nonumber\\
&=\frac{q^{\star}\left(v^{\star}-\gamma p^{\star}\right)}{\iota v^{\star 2}} \tilde{v}(x, t)  -\frac{\gamma p^{\star}}{\iota v^{\star}} \tilde{q}(x, t), \label{origin1}\\
\partial_t \tilde{v}(x, t)&-\left(\gamma p^{\star}-v^{\star}\right) \partial_x \tilde{v}(x, t)= \frac{\gamma p^{\star}-v^{\star}}{\iota v^{\star}} \tilde{v}(x, t)\nonumber\\
&-\frac{\gamma p^{\star}}{\iota q^{\star}} \tilde{q}(x, t),\label{origin2}
\end{align}
with boundary conditions 
\begin{align}
     \tilde{q}(0,t) = 0,
     \tilde{v}(L,t) = \frac{\tilde{q}(L,t)}{\rho^\star}+ U(t),
\end{align}
where $q(x,t)$ is the traffic flow, $v(x,t)$ denotes the traffic speed, defined in the spatial and time domain $(x,t) \in [0, L]\times [0, +\infty)$. The equilibrium state of the system as $(q^\star, v^\star)$, and the small deviations from the equilibrium points are defined as $\Tilde{q}(x,t) = q(x,t) - q^\star$, $\Tilde{v}(x,t) = v(x,t) - v^\star$. The traffic density is defined by $\rho(x,t) = \frac{q(x,t)}{v(x,t)}$. $v_f$ is the free-flow speed. The reaction time  $\iota$ denotes how long it takes for drivers' behavior to adapt to equilibrium speed. The traffic pressure is defined by $p = \frac{v_f}{\rho_m^\gamma}(\frac{q}{v})^\gamma$. $\gamma$ denotes the drivers' property which reflects their change of driving behavior to the increase of density.
Then we take the coordinate transformation to write the system in Riemann coordinates,
$\Bar{w} = \exp\left(\frac{x}{\iota v^\star}\right)\left(\Tilde{q} - q^\star\left(\frac{1}{v^\star} - \frac{1}{\gamma p^\star}\right)\right)\Tilde{v}$, $\Bar{v} = \left(\frac{q^\star}{\gamma p^\star}\right)\Tilde{v}$.
Then the control input $U(t)$, implemented by varying speed limit at the outlet of the road section, is given as $U_{\text{ARZ}}(t) = \mathsf{r} \Tilde{v}(L,t) +\mathsf{r} \int_0^L K^{vv}(L-\xi)\Tilde{v}(\xi,t) d\xi -\Tilde{q}(L,t)-\varrho_0 \int_0^L K^{vu}(L,\xi)\mathrm{e}^{\frac{\xi}{\iota v^\star}} \Tilde{v}(\xi,t)d\xi +\varphi_0 \int_0^L K^{vu}(L,\xi)\mathrm{e}^{\frac{\xi}{\iota v^\star}}\Tilde{q}(\xi,t) d\xi$, where $\mathsf{r} = q^\star(\frac{1}{v^\star} - \frac{1}{\gamma p^\star})$. The traffic density and speed would converge to the equilibrium density $\rho^\star$ and speed $v^\star$ with this control law at finite time. The ARZ system then transformed into boundary control model which aligns with \eqref{nom1}-\eqref{nom4} and the coefficients are $\lambda_0 = v^\star, \mu_0 = \gamma p^\star - v^\star, \sigma^+_0 = 0, \sigma^-_0 = -\frac{1}{\iota} \mathrm{e}^{-\frac{x}{\iota v\star}}, \varphi_0 = \frac{v^\star- \gamma p^\star}{v^\star}, \varrho_0 = \mathrm{e}^{-\frac{L}{\iota v^\star}}$.
The ARZ traffic system will be affected by the stochastic demand from upstream traffic flow. We consider the equilibrium density $\rho^\star(t)$ is stochastic and it follows the Markov process described in~\eqref{kolman}. Then the system parameters are all become stochastic due to the stochastic equilibrium density. Therefore, we can treat $(\lambda(t),\mu(t), \sigma^+(t),\sigma^-(t),\varphi(t),\varrho(t))$ as a whole Markov process. 
\subsection{Simulation configuration}
\begin{figure}
    \centering
    \subfloat[Probability]{\includegraphics[width=0.24\textwidth]{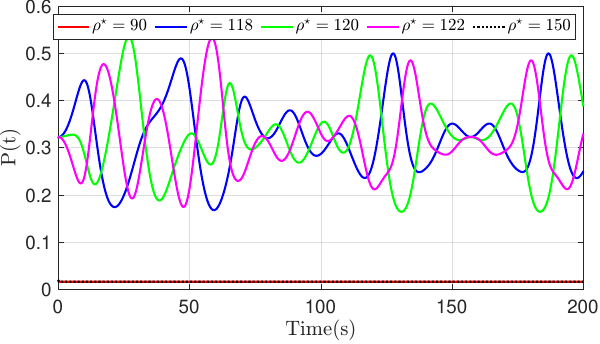}}
    \subfloat[States]{\includegraphics[width=0.24\textwidth]{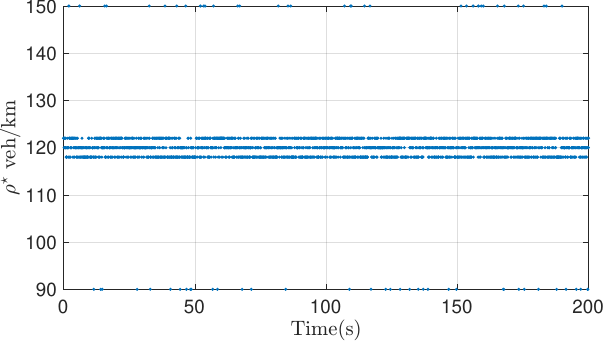}}
    \caption{The probability evolution and states reached in simulation time}
    \label{proandstates}
\end{figure}

We run the simulation on a $L=500\text{m}$ long road section and the simulation time is $T=200\text{s}$. The free-flow speed is $v_f = 144 \text{km/h}$, and the maximum density $\rho_m = 160 \text{veh/km}$, the nominal equilibrium states are choosen as $\rho^\star_0 = 120\text{veh/km}$, $v^\star = 36 \text{km/h}$. The reaction time $\iota = 60\text{s}$, and $\gamma=1$. We use sinusoidal inputs to denote stop-and-go traffic congestion on the road as $\rho(x,0) = \rho^\star + 0.1 \sin(\frac{3\pi x}{L})\rho^\star, v(x,0) = v^\star - 0.1 \sin(\frac{3 \pi x}{L}) v^\star$.
Considering the equilibrium density of the system as stochastic, we set five different settings of $\rho^\star$ as  {$(\rho^\star_1 = 90, \rho^\star_2 = 118, \rho^\star_3 = 120, \rho^\star_4 = 122, \rho^\star_5 = 150)$}. The initial probabilities are set as $(0.02,0.32,0.32,0.32,0.02)$, the transition rates $\tau_{ij}$ are defined as
\begin{align}
    \tau_{i j}(t)= \begin{cases}0, \quad \text { if } i=j \\ 20, \quad \text { if } i \in\{1,5\} \\ 10, \quad  \text { if } i \in\{2,3,4\}, j \in\{1,5\} \\ 10+20 \cos (0.01(i+5 j) t)^2, \text{others}  \end{cases}
\end{align}
Using the above settings, we numerically solve the Kolmogorov forward equation and get the results of the probability of each mode in the whole time period. The probability evolution is shown in Fig~\ref{proandstates}.
We adopted the DeepONet framework~\cite{lu2021learning} to train the neural operators in this paper. The DeepONet framework consists of brunch net and trunk net which can learn the different components of backstepping kernels. 
The input of the brunch net are the parameters of the ARZ system in the nominal mode, that is, $\lambda_0$, $\mu_0$, $\sigma^+_0$, $\sigma^-_0$, $\varphi_0$, and $\varrho_0$. The input of the truck net is chosen as the triangular domain grid.

 {To generate training dataset, we randomly sample the nominal equilibrium density $\rho^\star$ in the interval (90 veh/km, 130 veh/km) and calculate the corresponding nominal parameters related to the equilibrium density. And then using numerical method to solve backstepping kernel equations. Finally, we obtain the input parameters $\lambda_0, \mu_0, \sigma_0^+,\sigma_0^-, \varphi_0, \varrho_0$ and the exact backstepping kernels $K^{vu}(x,\xi), K^{vv}(x,\xi)$. The input-output pairs constitute the dataset. The dataset is divided into training dataset and testing dataset with the ratio of 9:1. All code and results necessary for reproducibility are provided openly at \url{https://github.com/curryzyang/NeuralOperator4RobustStabilization}}.

\subsection{Simulation results}
\begin{figure}[!tbp]
    \centering
    \subfloat[Traffic density]{\includegraphics[width=0.45\linewidth]{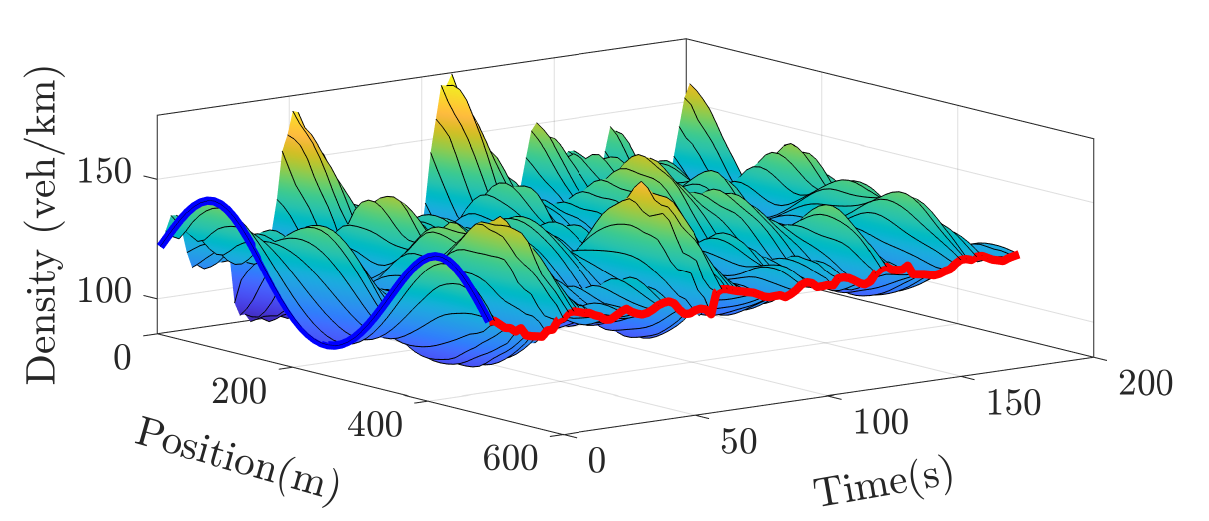}}
    \subfloat[Traffic speed]{\includegraphics[width=0.45\linewidth]{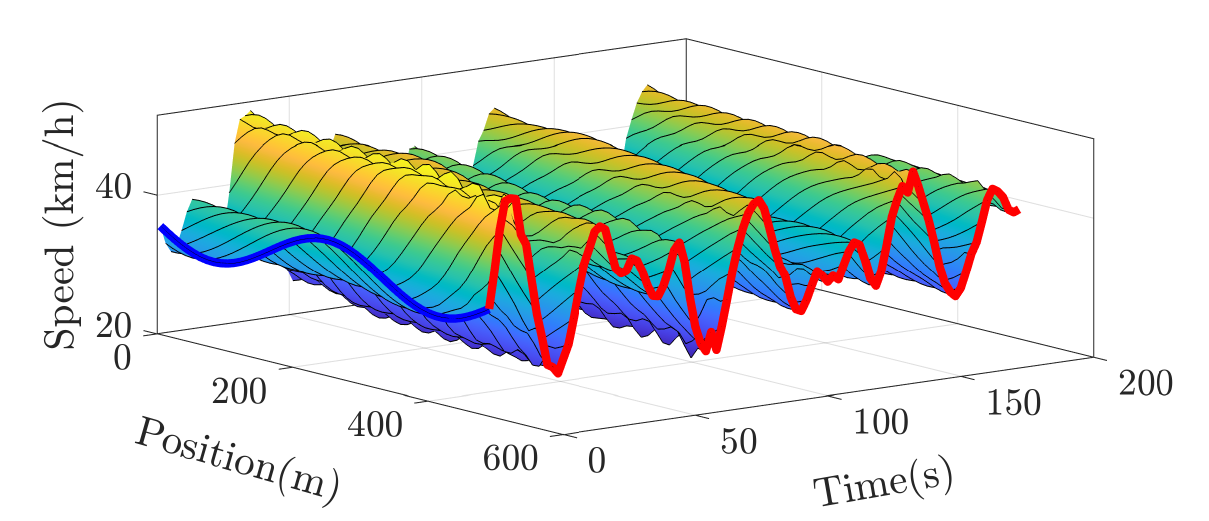}}
    \caption{The open-loop density and speed evolution}
    \label{olresults}
\end{figure}

\begin{figure}[!tbp]
    \centering
    \subfloat[Traffic density]{\includegraphics[width=0.45\linewidth]{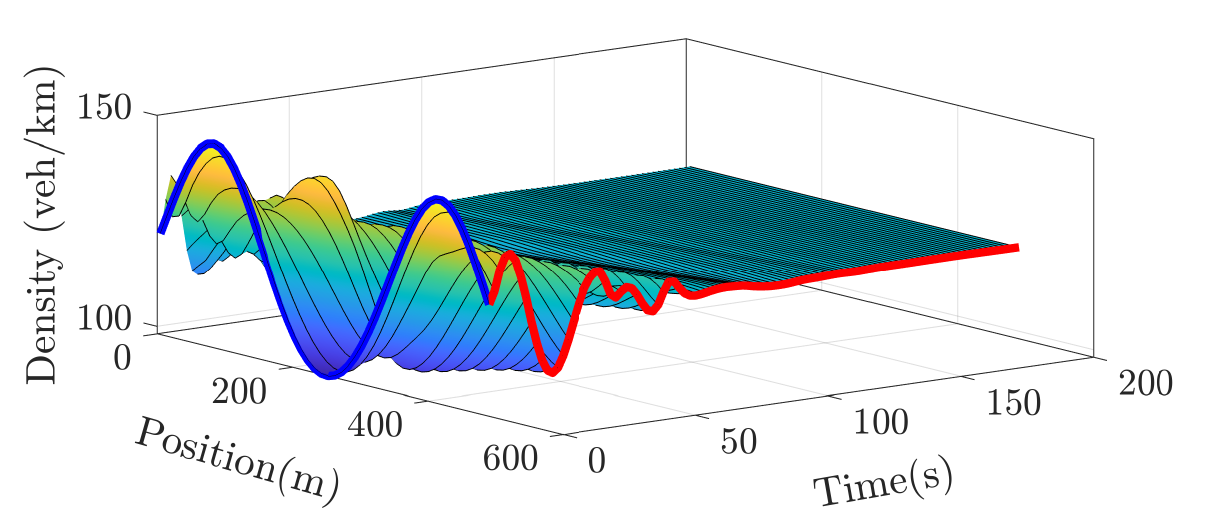}}
    \subfloat[Traffic speed]{\includegraphics[width=0.45\linewidth]{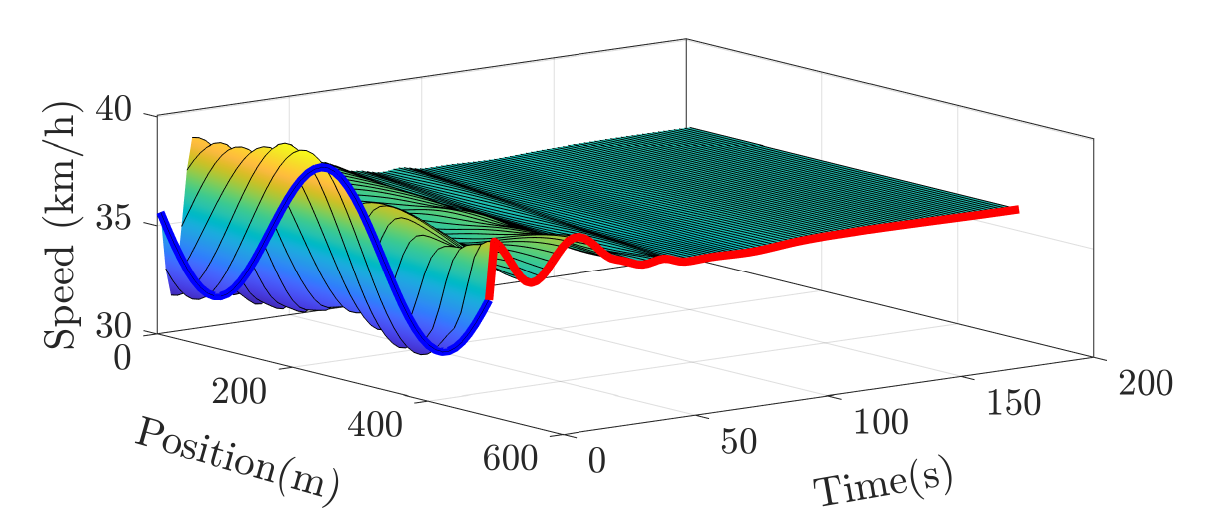}}
    \caption{The closed-loop density and speed evolution}
    \label{rhoandv}
\end{figure}

\begin{figure}[!tbp]
    \centering
    \subfloat[Traffic density]{\includegraphics[width=0.45\linewidth]{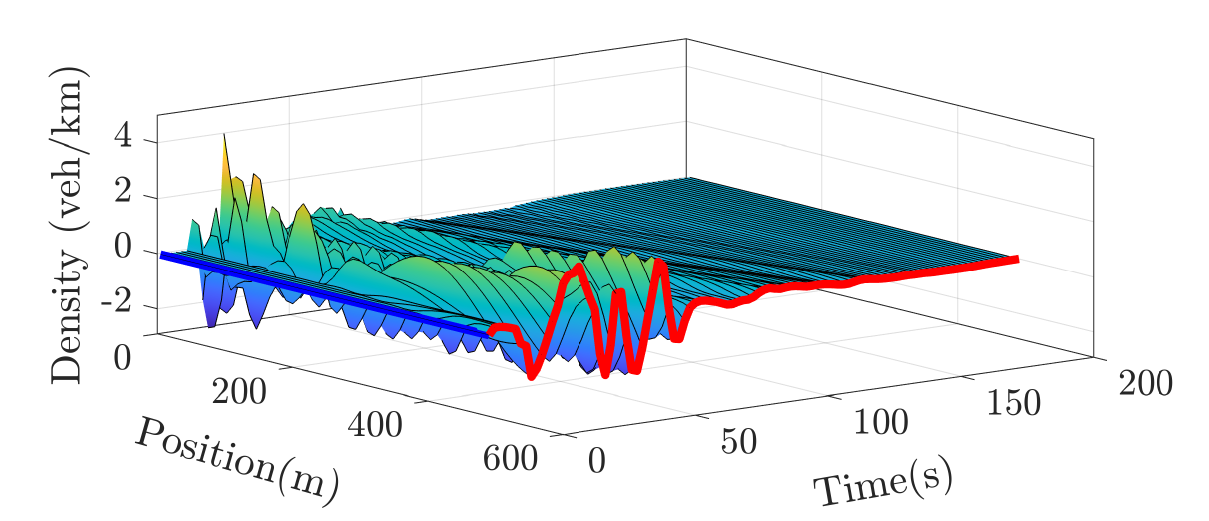}}
    \subfloat[Traffic speed]{\includegraphics[width=0.45\linewidth]{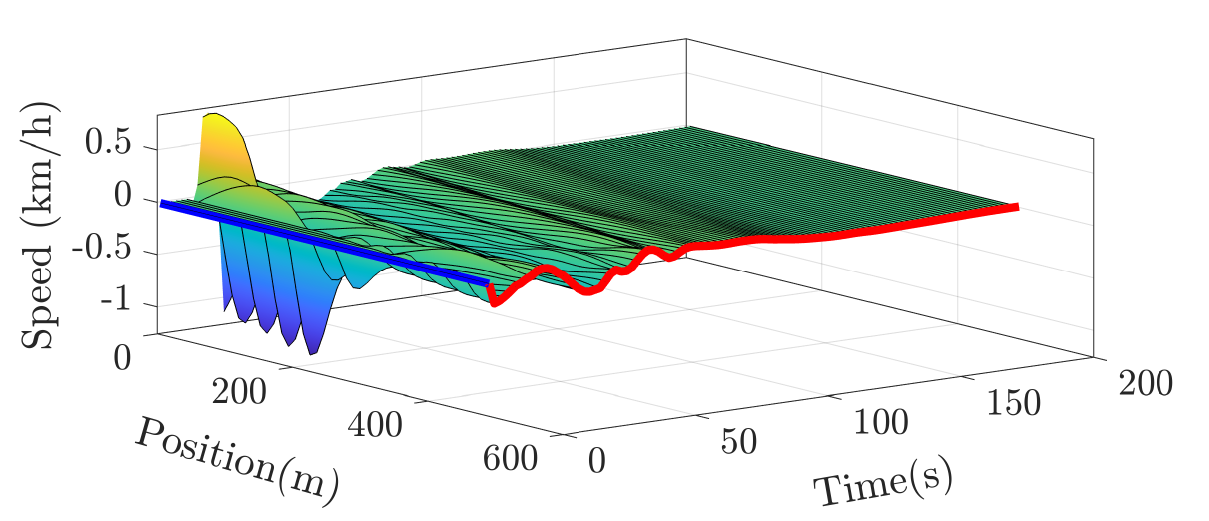}}
    \caption{The error of density and speed evolution}
    \label{errrhoandv}
\end{figure}

\begin{figure}[!tbp]
    \centering
    \subfloat[Control input]{\includegraphics[width=0.24\textwidth]{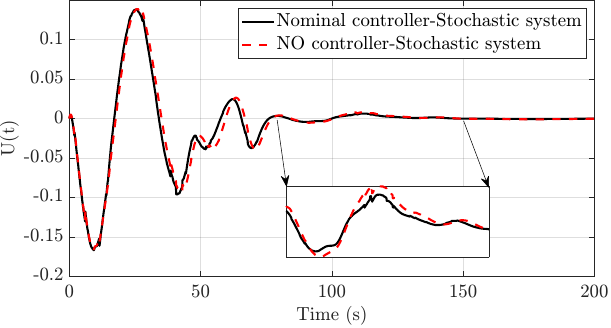}}
     \subfloat[States norm]{\includegraphics[width=0.24\textwidth]{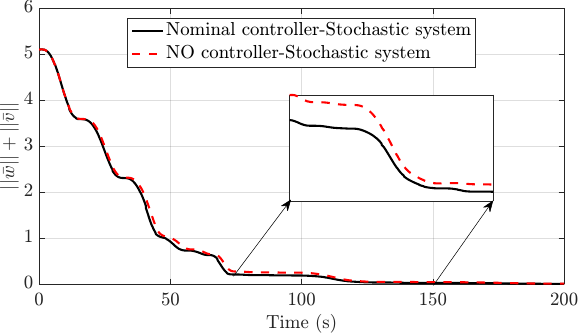}}
    \caption{The comparison of control input and state norm}
    \label{compareU}
\end{figure}

After the trained model is obtained, we test the model performance under the system with Markov-jumping parameters. The results of the open-loop density and speed evolution of the stochastic system are shown in Fig.~\ref{olresults}. The initial conditions and boundary conditions are denoted by a blue line and a red line, respectively. The open-loop density and speed of the stochastic system all oscillate during the simulation period. The results of the closed-loop with NO-approximated backstepping kernels are shown in Fig.~\ref{rhoandv}. It shows that the nominal controller with NO-approximated backstepping kernels successfully stabilizes the stochastic system. The traffic density and speed oscillations are removed after about 120s. We also compare the density and speed between the nominal controller and the nominal controller with NO-approximated backstepping kernels. The density and speed error is shown in Fig.~\ref{errrhoandv}. The density and speed errors are large and also oscillate at the initial stage of the simulation, then the error becomes small. The maximum density error is 4.04 veh/km while the maximum speed error is 1.31 km/h. 
The kernels obtained by NO and the numerical method lead to a different control input of the stochastic system. The control input for the ARZ system is obtained by integrating the production of kernels and traffic states of the stochastic system. The control input of the nominal controller with backstepping kernels and NO-approximated kernels are shown in Fig.~\ref{compareU}(a). Under different control inputs, we also compared the state norm of the traffic system, including the stochastic system with the nominal controller and the stochastic system with a nominal controller accompanied by NO-approximated kernels, as shown in Fig.~\ref{compareU}(b). 
From the comparison of control input and state norm, it is revealed that the nominal controller with NO-approximated kernels stabilizes the stochastic traffic system with small errors compared with the nominal controller with backstepping kernels. The statistical errors of kernels and traffic states are stated in Tab.~\ref{error_tab_rho_v_all}. After training, we take 100 trials to test the computation time of different methods. The average computation times for the backstepping and neural operator are $5.9899 \times 10^{-2}$s and  $1.7107\times 10^{-4}$s. It shows that the neural operator is 350$\times$ faster than the backstepping method.
\begin{table}[!tbp]
    \centering
    \caption{The error of backstepping kernels and traffic states}
    \begin{tabular}{c c c}
    \hline
       & \textbf{Max absolute error} & \textbf{Mean absolute error} \\
    \hline
    $K^w$ & $5.9604 \times 10^{-5}$ & $3.0066\times 10^{-5}$ \\
    
    $K^v$ & $5.9519\times 10^{-5}$ & $2.2160\times 10^{-5}$\\
    
    $\rho$ & $4.0456 $ & $0.2450$  \\
    
    $v$ & $1.3119$ & $0.0476$ \\
    \hline
    \end{tabular}
    \label{error_tab_rho_v_all}
\end{table}

\section{Conclusions}\label{sec6}
In this paper, we investigated the mean-square exponential stability of the hyperbolic PDEs with Markov-jumping parameters under the nominal controller constructed with NO-approximated backstepping kernels. We use neural operators to approximate the nominal backstepping kernel gains. The Markov-jumping hyperblic PDE system with the NO-approximated control law achieves mean-square exponential stability, provided the stochastic parameters are close to the nominal parameters. The theoretical result is obtained through the Lyapunov analysis and it was applied to freeway traffic congestion mitigation. The simulation results demonstrate that the neural operator stabilizes the stochastic system with a 350$\times$ computation speed faster than the numerical method. Future work would focus on the observer design for the stochastic PDE system and extension to $n+m$ systems.

\begin{ack}                               
This work was supported by the National Natural Science Foundation of China No.62203131 and Guangzhou Municipal Education Bureau University Project 2024312102.  
\end{ack}

\bibliographystyle{abbrv}        
\bibliography{reference}           

\end{document}